\documentclass[%
 aip,
 rsi,
 amsmath,
 amssymb,
 reprint,%
]{revtex4-2}

\usepackage{graphicx}
\usepackage{dcolumn}
\usepackage{bm}

\usepackage[utf8]{inputenc}
\usepackage[T1]{fontenc}
\usepackage{mathptmx}
\usepackage{etoolbox}

\usepackage{physunits}
\usepackage{siunitx}

\usepackage{hyperref}

\hypersetup{
    colorlinks=true,
    linkcolor=blue,
    citecolor=blue,
    filecolor=blue,
    urlcolor=blue
}

\begin{document}

\title{Fast, optimal readout of a tethered optomechanical pressure sensor} 

\author{Stephen R. Chen}
\altaffiliation{Present address: Massachusetts Institute of Technology, Cambridge, MA 02139, USA}
\author{Yiliang Bao}
\author{John R. Lawall}
\author{Jason J. Gorman}
\author{Daniel S. Barker}
\email{daniel.barker@nist.gov}
\affiliation{National Institute of Standards and Technology, Gaithersburg, MD 20899, USA}

\date{\today}

\begin{abstract}
We demonstrate a fast readout method for tethered optomechanical pressure sensors.
The fast readout method allows mechanical ring-down measurement rates that exceed the total mechanical damping rate \(\Gamma_t\).
We model the uncertainty of our ring-down measurements with the Cram\'er-Rao bound, including the effect of thermomechanical noise.
Both our data and the Cram\'er-Rao bound indicate that measurement rates on the order of the mechanical damping rate produce the lowest uncertainty.
At low pressures, using the optimal measurement rate allows \(10\times\) faster measurements or \(3\times\) lower uncertainty compared to allowing our sensor to ring down to the thermomechanical noise floor.
Our results allow tethered optomechanical pressure sensors to operate at rates comparable to commercially available pressure gauges, removing an obstacle to adoption of optomechanical pressure sensors in industrial settings. 
\end{abstract}


\maketitle 

\section{Introduction}
\label{sec:intro}

There has been significant recent interest in developing pressure gauges based on optomechanical systems.\cite{Scherschligt2018, Blakemore2020, salimi2024, reinhardt2024, Liu2024, barker2024, Green2025, Gajewski2025, Reinhardt2026, tseng2026}
Optomechanical pressure sensors determine the pressure rise above background via gas-induced resonance frequency shifts\cite{salimi2024,reinhardt2024} or mechanical damping.\cite{lubbe2011, Blakemore2020, reinhardt2024, Liu2024, Green2025, Reinhardt2026}
Optomechanical pressure sensors that deduce the pressure from measurements of the total mechanical damping rate \(\Gamma_t\) have been shown to operate over a wide pressure range and be accurate to approximately \(10~\si{\percent}\).\cite{reinhardt2024, Liu2024, Green2025, Reinhardt2026}
Mechanical-damping-based optomechanical pressure sensors have a slow pressure measurement rate because low uncertainty determination of the mechanical damping rate typically requires measurement time substantially longer than \(1/\Gamma_t\).\cite{lubbe2011, Liu2024, reinhardt2024, Green2025} 
For tethered optomechanical pressure sensors, the demonstrated pressure measurement time is on the order of minutes in the high-vacuum range (\(10^{-6}~\si{\pascal}\) to \(0.1~\si{\pascal}\)).\cite{lubbe2011, reinhardt2024, Green2025}
The long measurement times for damping-based optomechanical pressure sensors is a significant obstacle to their adoption outside of research environments.

Here, we demonstrate a fast readout scheme for tethered, mechanical-damping-based optomechanical pressure sensors.
Inspired by fast readout methods for magnetically and optically levitated spinning rotor gauges,\cite{Fremerey1985,Blakemore2020} we use a phase-locked loop to coherently drive our optomechanical pressure sensor.
Our method allows repeatable re-excitation of the sensor's mechanical motion without waiting for it to damp to the thermomechanical noise floor.
As a result, we can perform ring-down measurements at rates exceeding \(\Gamma_t\).

We use our method (detailed in Sec.~\ref{sec:procedure}) to study the competition between fast measurement rates and low statistical uncertainty in a given total measurement time.
To understand our measurements, we derive the Cram\'er-Rao bound for the mechanical ring-down, including the effect of thermomechanical noise, in Sec.~\ref{sec:cr}.
The Cram\'er-Rao bound indicates that the minimum statistical uncertainty is reached when the ring-down is measured for a duration much shorter than those used in prior studies of tethered optomechanical pressure sensors.\cite{lubbe2011, reinhardt2024, Green2025}
In Sec.~\ref{sec:opt}, we find that our experimentally achieved uncertainties are well-described by the Cram\'er-Rao bound.
By measuring each ring-down for the optimal time, we can substantially improve the pressure resolution or measurement rate of our optomechanical pressure sensors.
We summarize our results and provide outlook in Sec.~\ref{sec:con}.

\section{Fast Readout Method}
\label{sec:procedure}

\begin{figure}
    \includegraphics[width=\columnwidth]{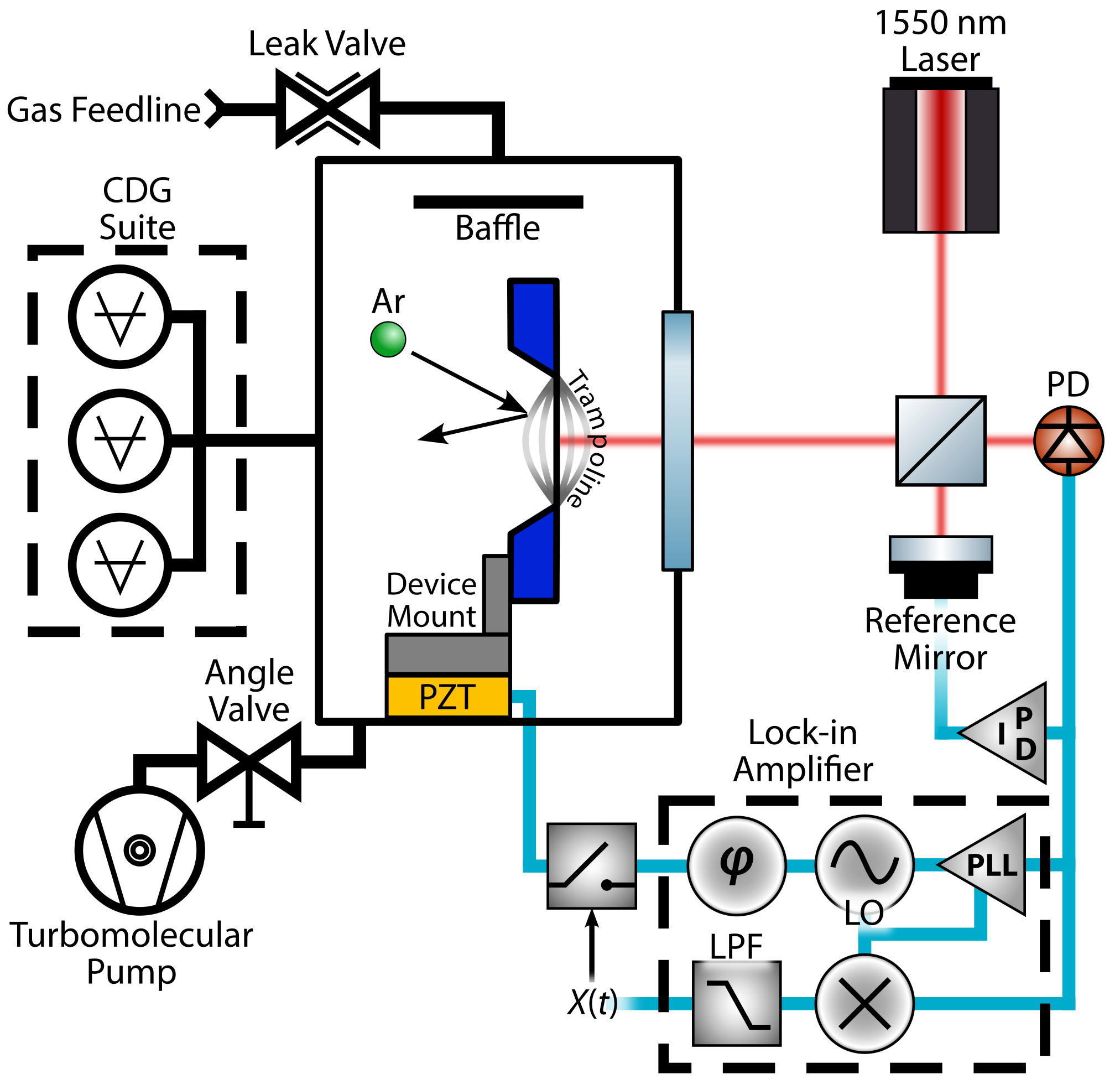}
    \caption{\label{fig:apparatus}
    Schematic of the experimental apparatus.
    Abbreviations -- CDG: capacitance diaphragm gauge, LO: local oscillator, LPF: low-pass filter, PD: photodiode, PID: proportional-integral-differential controller, PLL: phase-locked loop, PZT: piezoelectric transducer.
    }
\end{figure}

\begin{figure*}
    \includegraphics[width=\textwidth]{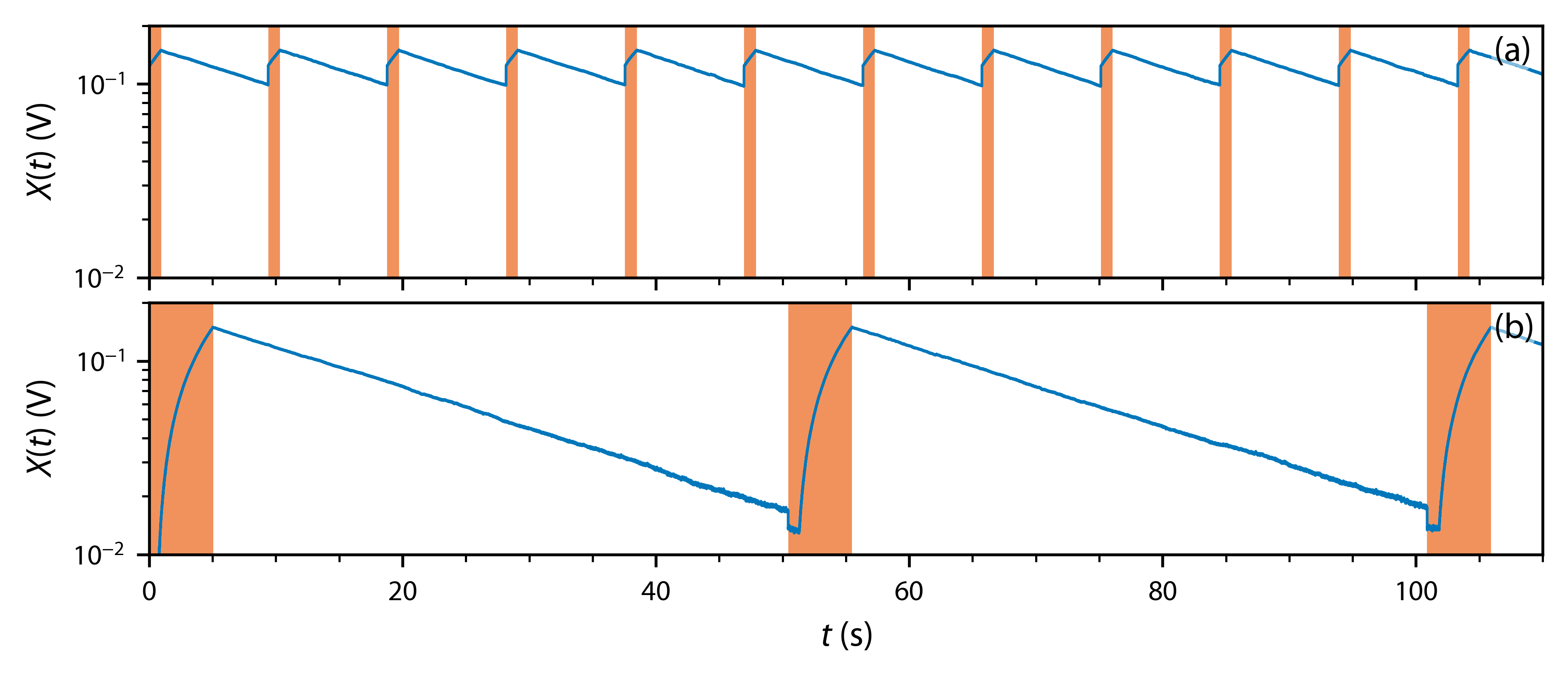}
    \caption{\label{fig:example}
    Example mechanical ring-down signals.
    Blue lines in (a) and (b) show the in-phase quadrature of the mechanical motion \(X(t)\) for measurement times of \(t_{RD}\approx 0.8/\Gamma_t\) and \(t_{RD}\approx 4.4/\Gamma_t\), respectively.
    Orange rectangles cover data taken during re-excitation of the trampoline.
    }
\end{figure*}

The main features of our apparatus, which is shown in Fig.~\ref{fig:apparatus}, have been previously described in Ref.~\onlinecite{Green2025}.
Briefly, we measure the mechanical damping response to gas pressure of an optomechanical trampoline resonator using a homodyne Michelson interferometer.\cite{Norte2016, Reinhardt2016}
The trampoline is mounted on a piezoelectric shaker in a vacuum chamber that is evacuated to a base pressure of approximately \(2\times10^{-5}~\si{\pascal}\) with the residual gas mostly consisting of water vapor.
We introduce \(99.999~\si{\percent}\) pure Argon test gas (UHP grade) into the vacuum chamber using a variable leak valve.
The leak valve can set test gas pressures from approximately \(10^{-5}~\si{\pascal}\) to approximately \(1~\si{\pascal}\) and is stable for approximately \(1000~\si{\second}\),\cite{Green2025} which is sufficient for the measurements in Sec.~\ref{sec:opt}.
A suite of calibrated capacitance diaphragm gauges measures the test gas pressure.
A baffle located in front of the gas inlet ensures that the test gas thermalizes with the vacuum chamber walls before interacting with either the optomechanical pressure sensor or the capacitance diaphragm gauges.
Two calibrated platinum resistance thermometers measure the vacuum chamber temperature to determine the test gas temperature.

A lock-in amplifier and phase-locked loop (PLL) monitor the voltage signal from the Michelson interferometer (see Fig.~\ref{fig:apparatus}).
The PLL continuously phase locks the lock-in amplifier's local oscillator (LO) to the mechanical motion of the trampoline.
The lock-in amplifier's in-phase quadrature output \(X(t)\) is therefore proportional to the amplitude of the trampoline's motion \(A(t)\) at time \(t\).
We excite the mechanical motion of the trampoline by sending a phase-shifted copy of the LO through a fast switch to the piezoelectric shaker.
The lock-in amplifier turns off the switch when \(X(t)\) reaches a preset threshold voltage.
The mechanical motion then rings down according to
\begin{equation}
    \label{eq:quad}
    X(t) = V_{\rm exc} e^{-\Gamma_t (t-t_s)/2}+V_0,
\end{equation}
where the ring-down begins at time \(t_s\), \(V_{\rm exc}\) is the initial excitation amplitude, and \(V_0\) is a stable background voltage.
The background \(V_0\) arises due to the trampoline's thermomechanical noise and the finite isolation of the switch.
The lock-in amplifier records the mechanical ring-down for a programmable time \(t_{RD}\) before turning the switch on to excite another ring-down.
In this study, we use ring-down times from \(t_{RD}\approx 100~\si{\milli\second}\) to \(t_{RD}\approx 120~\si{\second}\), but shorter or longer ring-down times are possible.

The lock-in amplifier and PLL allow us to repeatably re-excite the trampoline's motion without waiting for it to ring down to the measurement noise floor.
As a result, we can substantially increase the ring-down measurement rate.
Figure~\ref{fig:example} shows examples of repetitive ring-down measurements at base pressure recorded with \(t_{RD}\approx 8.5~\si{\second}\approx 0.8/\Gamma_t\) (Fig.~\ref{fig:example}(a)) and \(t_{RD}\approx 45~\si{\second}\approx 4.4/\Gamma_t\) (Fig.~\ref{fig:example}(b)).
Orange regions in Fig.~\ref{fig:example} indicate data acquired during excitation of the trampoline.
Our ring-down measurements are not continuous (as indicated by the discontinuities in the data at the left edge of each orange region in Fig.~\ref{fig:example}), so the orange regions only partially account for the measurement dead time \(t_d\).
The dead time is limited by the time required to re-excite the trampoline between ring-downs and by the acquisition of data during the excitation.
We find that
\begin{equation}
    \label{eq:dead}
    t_d\approx t_{d,0} + a_d t_{RD} + t_{\rm drv}(1-e^{-\Gamma_t t_{RD}/2}),
\end{equation}
where \(t_{d,0}\lesssim 50~\si{\milli\second}\) is the minimum dead time of the lock-in amplifier (which depends on the data sampling rate), \(a_d \approx 0.11\) arises from the programmed acquisition of excitation data (orange regions in Fig.~\ref{fig:example}), and \(t_{\rm drv}\) represents the time to excite the trampoline to the threshold voltage from the thermomechanical noise floor.
For the range of \(t_{RD}\) that we explore, the measurement duty cycle \(t_{RD}/(t_{RD}+t_d)\) varies from approximately \(70~\si{\percent}\) to approximately \(80~\si{\percent}\) with higher duty cycles occurring at higher \(t_{RD}\).

We extract \(\Gamma_t\) for each ring-down from exponential fits to Eq.~\eqref{eq:quad}, excluding data collected during device excitation.
Because most ring-downs terminate at \(X(t)\gg V_0\), we determine \(V_0\) from an independent \(X(t)\) measurement with the switch off and fix \(V_0\) in our exponential fits.  
The pressure \(P\propto\Gamma_t-\Gamma_0\) can be computed following Refs.~\onlinecite{Cavalleri2010, Martinetz2018, Green2025}, where \(\Gamma_0\) is the intrinsic (\textit{i.e.} pressure-independent) damping rate of the trampoline and we take \(\Gamma_0\) equal to the total mechanical damping rate measured at base pressure.


The ring-down measurements in Fig.~\ref{fig:example} suggest a question: what is the trade-off between the standard uncertainty in the measured mechanical damping rate \(u(\Gamma_t)\) and the measurement rate (proportional to \(1/t_{RD}\)) at fixed total measurement time \(\tau\)?
That is to say, what is the optimal number of individual ring-downs to measure in a given total measurement time~$\tau$?
The standard uncertainty in a set of \(M\) individual ring-down measurements is
\begin{equation}
    \label{eq:stand_unc}
    u(\Gamma_t)=\frac{s(\Gamma_t)}{\sqrt{M}}=\frac{s(\Gamma_t)}{\sqrt{\tau/(t_{RD}+t_d)}},
\end{equation}
where \(s(\Gamma_t)\) is the standard deviation of the repeated ring-down measurements and we note that \(s(\Gamma_t)\) is a function of \(t_{RD}\) as we discuss below.
For \(t_{RD}\ll1/\Gamma_t\), the change in motion amplitude \(\Delta X(t_{RD})=V_{\rm exc}(1-e^{-\Gamma_t t_{RD}/2})\) will be comparable to the voltage noise in \(X(t)\); leading to large individual measurement uncertainty \(s(\Gamma_t)\) that is not compensated by the large number of measurements $M$.
For \(t_{RD}\gg1/\Gamma_t\), data collected late in the ring-down will have poor signal-to-noise ratio (see Fig.~\ref{fig:example}(b)) and will not efficiently reduce \(s(\Gamma_t)\).
We therefore expect that there is an optimal \(t_{RD}\) on the order of \(1/\Gamma_t\) that minimizes the standard uncertainty \(u(\Gamma_t)\) in the set of \(M\) individual measurements  at fixed \(\tau\).

\begin{figure}[b]
    \includegraphics[width=\columnwidth]{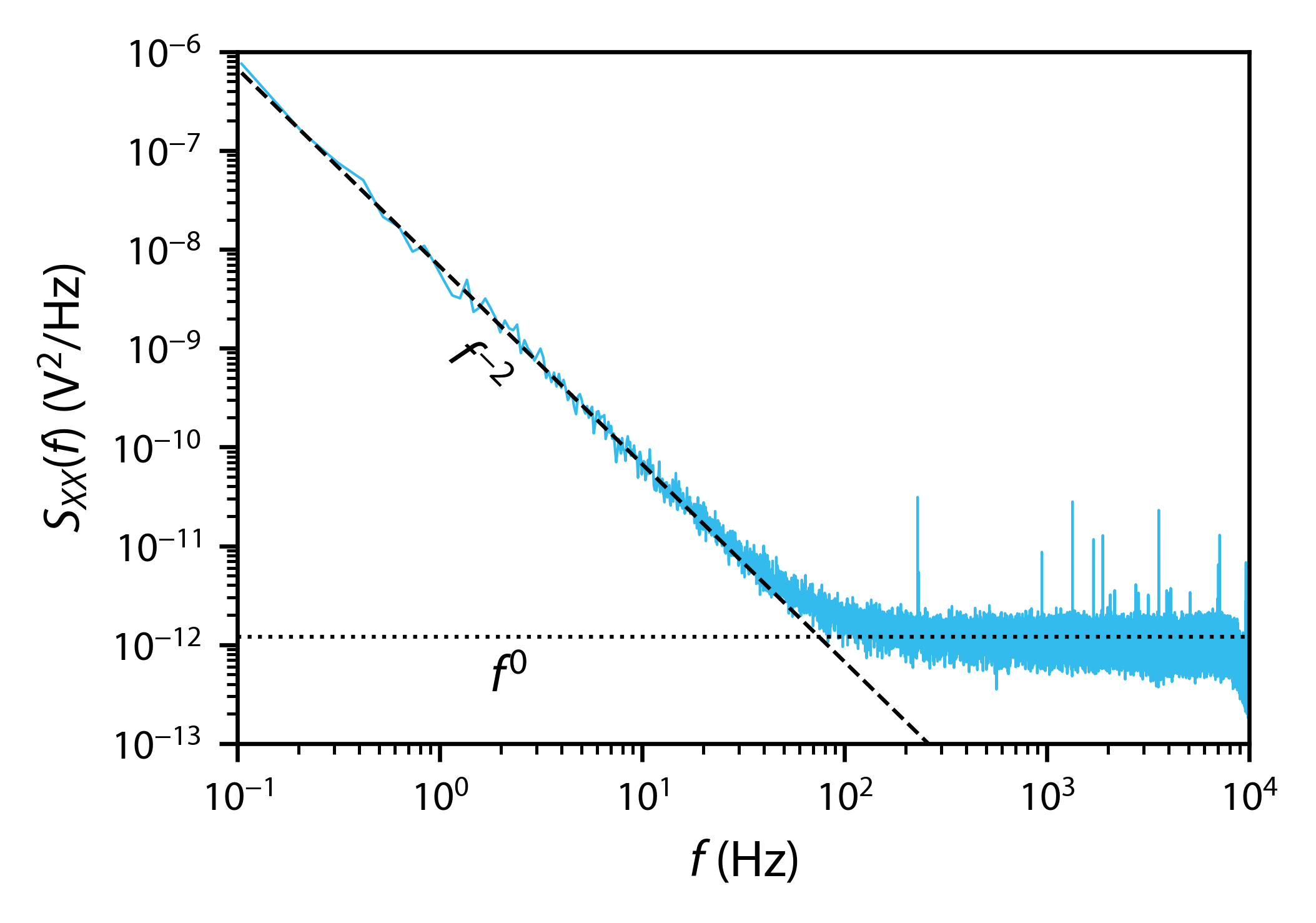}
    \caption{\label{fig:psd}
    Example power spectral density of the noise in the in-phase quadrature \(S_{XX}(f)\) at base pressure (blue).
    The black dashed and dotted lines show a \(f^{-2}\) and \(f^0\) frequency dependence, respectively.
    The roll-off of the power spectral density near \(10~\si{\kilo\hertz}\) arises from the cut-off frequency of the lock-in amplifier's low-pass filter.
    }
\end{figure}

The uncertainties \(s(\Gamma_t)\) of the individual measurements in Eq.~\eqref{eq:stand_unc} depend on the measurement noise of the data points comprising each ring-down.
The nature of the measurement noise is captured by the power spectral density of the in-phase quadrature \(S_{XX}(f)\), shown in Figure~\ref{fig:psd}, where \(f\) is the frequency.
The power spectral density has a frequency dependence of \(f^{-2}\) over several decades at low frequency, arising from thermomechanical noise.\cite{Fong2012}
We have verified that the \(f^{-2}\) noise is absent if the trampoline is replaced with a mirror.
At higher frequencies, the thermomechanical noise is overwhelmed by other noise sources, such as amplifier noise and demodulated laser noise, with a nominally white (\(f^0\)) frequency dependence.
Integration of the power spectral density in Figure~\ref{fig:psd} reveals that in the frequency range shown, the mean-square thermal noise is approximately~5 times larger than the mean-square white noise.
The white noise is uncorrelated, while the thermal noise is correlated over a time of the order of $2/\Gamma_t$ (see Appendix~\ref{sec:psd}).
The correlated noise creates the small wiggles visible near the end of each ring-down in Fig.~\ref{fig:example}(b).
Longer individual ring-down measurement times \(t_{RD}\) entail signal fluctuations arising from integrating the noise in Figure~\ref{fig:psd} to progressively lower frequencies, rapidly increasing the amount of noise.

\section{Cram\'er-Rao Bound}
\label{sec:cr}

We quantitatively study the effects of thermomechanical noise and white noise on the standard uncertainty \(u(\Gamma_t)\) using the Cram\'er-Rao bound.
The Cram\'er-Rao bound is the minimum standard deviation \(s_{\rm min}(\theta_\alpha)\) of a parameter \(\theta_\alpha\) that can be achieved for a given measurement protocol and noise level, where \(\alpha\) indexes the individual parameters in the parameter set \(\vec{\theta}\).\cite{Rao1945, Cramer1946, Jones1996}
For the mechanical ring-down described by Eq.~\eqref{eq:quad}, the parameters are \(\theta_\alpha\in\{V_{\rm exc}, \Gamma_t\}\), and we wish to calculate the Cram\'er-Rao bound for \(s(\Gamma_t)\).
Derivation of the Cram\'er-Rao bound requires a full statistical model of the set of \(X(t)\) measurements that constitute a ring-down (\textit{i.e.} the model must include the mean, covariance, and distribution function of the measurements).
We model the \(X(t)\) measurements as a set of \(N\) Gaussian-distributed random variables \(\vec{X}\) (for our measurements, \(10^4 \lesssim N \lesssim 10^7\)).
The mean value of each random variable \(X_j\in\vec{X}\) is given by
\begin{equation}
    \label{eq:rand}
    \overline{X}_j = V_{\rm exc} e^{-\Gamma_t jt_0/2},
\end{equation}
where \(1/t_0\) is the sampling rate, \(j\) is an integer, and \(V_0\) does not appear because it is fixed in our exponential fits to the data (see Sec.~\ref{sec:procedure}).
The elements of the covariance matrix for the \(X_j\) are given by (see Appendix~\ref{sec:covar})
\begin{equation}
    \label{eq:cov}
    C_{jk} = \sigma^2_w\delta_{jk}+\sigma^2_{th}e^{-\Gamma_t t_0|k-j|/2},
\end{equation}
where \(k\) is an integer, \(\sigma^2_w\) is the variance of the white noise, \(\sigma^2_{th}\) is the variance of the thermomechanical noise, and \(\delta_{jk}\) is the Kronecker delta.

The Cram\'er-Rao bound for \(\theta_\alpha\in\{V_{\rm exc}, \Gamma_t\}\) is
\begin{equation}
    \label{eq:bound}
    s_{\rm min}(\theta_\alpha) = \sqrt{(\mathcal{I}^{-1})_{\alpha\alpha}},
\end{equation}
where \(\mathcal{I}\) is the Fisher information matrix.\cite{Jones1996}
The elements of the Fisher matrix are\cite{Malag2015}
\begin{equation}
    \label{eq:fisher}
    \begin{split}
    \mathcal{I}_{\alpha\beta} & = \sum^{N-1}_{j=0}\sum^{N-1}_{k=0}\frac{\partial \overline{X}_j}{\partial\theta_\alpha}(C^{-1})_{jk}\frac{\partial \overline{X}_k}{\partial\theta_\beta} \\
    & + \frac{1}{2}\sum^{N-1}_{j=0}\sum^{N-1}_{k=0}\sum^{N-1}_{m=0}\sum^{N-1}_{n=0}(C^{-1})_{jk}\frac{\partial C_{km}}{\partial\theta_\alpha}(C^{-1})_{mn}\frac{\partial C_{nj}}{\partial\theta_\beta},
    \end{split}
\end{equation}
where \(m\) is an integer, \(n\) is an integer, and \(\beta\) indexes the parameters in \(\vec{\theta}\).
When \(\sigma_w = 0\) or \(\sigma_{th} = 0\), there are simple expressions for \(C^{-1}\) and we can calculate \(s_{\rm min}(\theta_\alpha)\) analytically.
When \(\sigma^2_{th}=0\) and \(t_d=0\), the elements of the Fisher matrix (Eq.~\eqref{eq:fisher}) are geometric series and
\begin{widetext}
    \begin{equation}
        \label{eq:white}
        s_{\rm min}(\Gamma_t) = \frac{2\sigma_w}{V_{\rm exc}}\sqrt{\frac{(e^{-\Gamma_t t_0}-1)^3(e^{-\Gamma_t(t_{RD}+t_0)}-1)}{t_0^2 e^{-\Gamma_t t_0}(1+e^{-2\Gamma_t(t_{RD}+t_0)})-e^{-\Gamma_t(t_{RD}+t_0)}(2t_0^2 e^{-\Gamma_t t_0}+(t_{RD}+t_0)^2(e^{-\Gamma_t t_0}-1)^2)}}.
    \end{equation}
When \(\sigma_w^2=0\) and \(t_d=0\), the Fisher information matrix is diagonal and
    \begin{equation}
        \label{eq:thermomech}
        s_{\rm min}(\Gamma_t) = \frac{2\sigma_{th}}{t_0}\frac{1-e^{-\Gamma_t t_0}}{\sqrt{V_{\rm exc}^2e^{-\Gamma_t t_0}(1-e^{-\Gamma_t t_{RD}})+\sigma_{th}^2e^{-\Gamma_t t_0}(e^{-\Gamma_t t_0}+1)t_{RD}/t_0}}.
    \end{equation}
\end{widetext}
When \(\sigma_w^2>0\) and \(\sigma^2_{th}>0\), the sums in Eq.~\eqref{eq:fisher} do not condense into simple analytic expressions when the number of \(X(t)\) measurements \(N>2\) and we therefore calculate the Cram\'er-Rao bound (Eq.~\eqref{eq:bound}) numerically.
However, when \(N=2\), the Cram\'er-Rao bound for \(s(\Gamma_t)\) can be computed analytically for any \(\sigma_w^2\) and \(\sigma_{th}^2\).
When \(V_{\rm exc}\gg\sigma^2_{th}\), the result is
\begin{equation}
    \label{eq:2point}
    s_{\rm min}(\Gamma_t) \approx \frac{2}{V_{\rm exc}t_{RD}}\sqrt{\sigma^2_w(1+e^{\Gamma_t t_{RD}})+\sigma^2_{th}(e^{\Gamma_t t_{RD}}-1)},
\end{equation}
which becomes exact and agrees with Ref.~\onlinecite{Jones1996} when \(\sigma^2_{th}=0\).

\begin{figure}
    \includegraphics[width=\columnwidth]{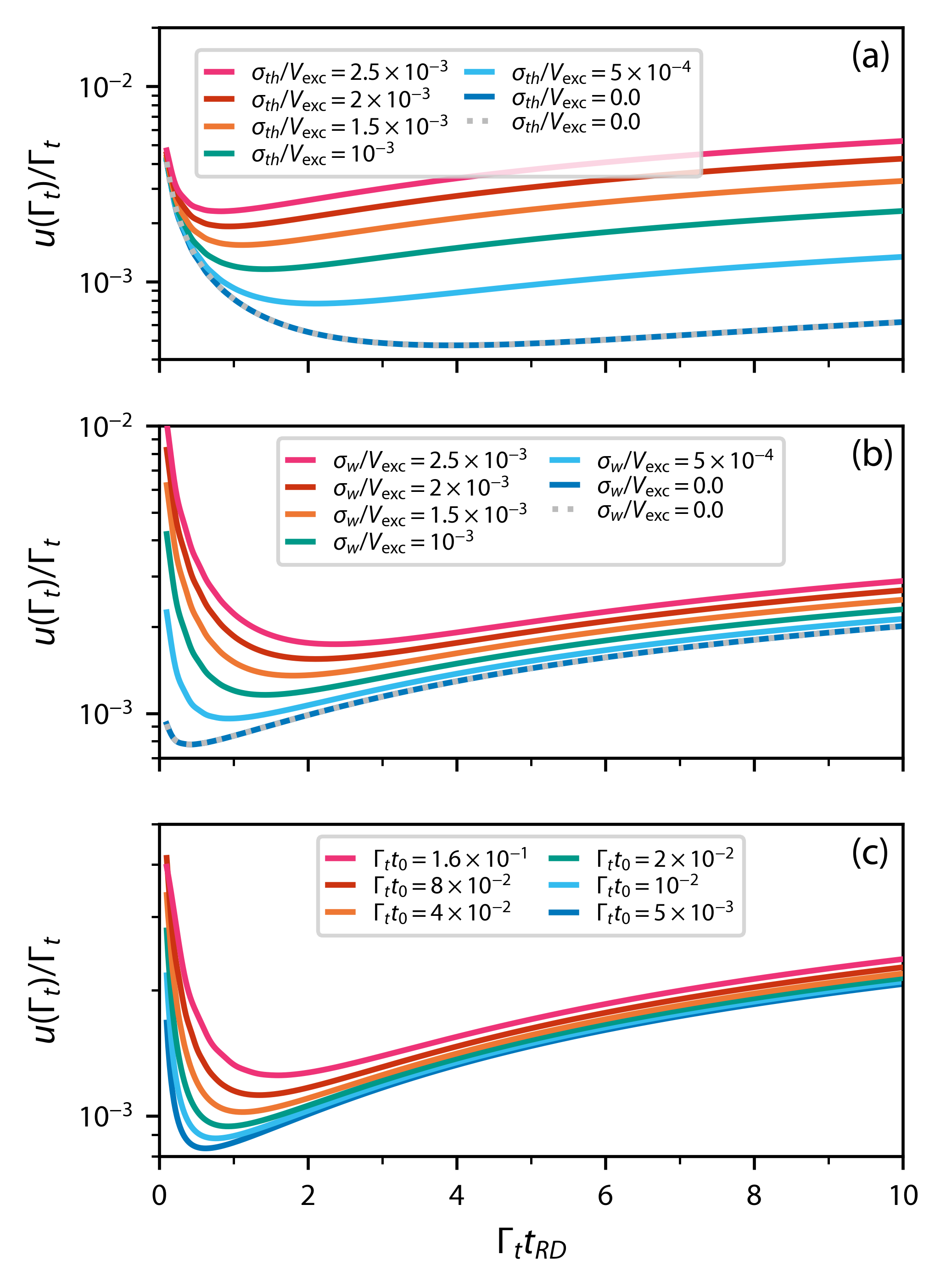}
    \caption{\label{fig:unc_thry}
    Cram\'er-Rao bound for the fractional standard uncertainty in \(\Gamma_t\) as function of the individual ring-down measurement time \(t_{RD}\) at fixed total measurement time \(\tau = 10/\Gamma_t\).
    In all subplots, solid curves show \(u(\Gamma_t)/\Gamma_t\) calculated numerically using Eq.~\eqref{eq:bound} with \(V_{\rm exc}=100~\si{\milli\volt}\).
    Subplot (a) shows the effect of varying the thermal noise \(\sigma_{th}\) for fixed white noise \(\sigma_w/V_{\rm exc}=1\times10^{-3}\) and sampling time \(\Gamma_t t_0=0.1\), with \(\sigma_{th}/V_{\rm exc}\) indicated by the legend.
    The gray, dotted curve shows the analytic result for \(u(\Gamma_t)/\Gamma_t\) derived from Eq.~\eqref{eq:white}.
    Subplot (b) shows the effect of varying the white noise \(\sigma_{w}\) for fixed thermal noise \(\sigma_{th}/V_{\rm exc}=1\times10^{-3}\) and sampling time \(\Gamma_t t_0=0.1\), with \(\sigma_w/V_{\rm exc}\) indicated by the legend.
    The gray, dotted curve shows the analytic result for \(u(\Gamma_t)/\Gamma_t\) derived from Eq.~\eqref{eq:thermomech}.
    Subplot (c) shows the effect of varying the sampling time \(t_0\) for fixed white noise \(\sigma_w/V_{\rm exc}=1\times10^{-3}\) and thermal noise \(\sigma_{th}/V_{\rm exc}=1\times10^{-3}\), with \(\Gamma_t t_0\) indicated by the legend.
    }
\end{figure}

We are interested in the dependence of \(u(\Gamma_t)\) on the thermal noise variance, white noise variance, and sampling rate.
To find this dependence, we calculate \(u(\Gamma_t)\) with Eq.~\eqref{eq:stand_unc}, replacing the experimental \(s(\Gamma_t)\) with the Cram\'er-Rao bound \(s_{\rm min}(\Gamma_t)\) from Eq.~\eqref{eq:bound}, Eq.~\eqref{eq:white}, or Eq.~\eqref{eq:thermomech}.
We express the noise in units of the excitation amplitude \(V_{\rm exc}\) and the time in units of \(1/\Gamma_t\).
Figure~\ref{fig:unc_thry} shows the resulting fractional uncertainty \(u(\Gamma_t)/\Gamma_t\) at fixed total measurement time \(\tau\) and dead time \(t_d=0\) with variable  \(\sigma_{th}/V_{\rm exc}\) (Fig.~\ref{fig:unc_thry}(a)), \(\sigma_{w}/V_{\rm exc}\) (Fig.~\ref{fig:unc_thry}(b)), or \(\Gamma_t t_0\) (Fig.~\ref{fig:unc_thry}(c)).
The Cram\'er-Rao bounds exhibit minima at an optimal individual ring-down time \(t_{RD}\) that is much shorter than \(\Gamma_t t_{RD}\approx 10\) employed in Refs.~\onlinecite{reinhardt2024, Green2025}.
Consistent with our expectations, increasing the thermomechanical noise \(\sigma_{th}/V_{\rm exc}\) decreases the optimal \(\Gamma_t t_{RD}\) (see Fig.~\ref{fig:unc_thry}(a)), while increasing the white noise \(\sigma_w/V_{\rm exc}\) increases the optimal \(\Gamma_t t_{RD}\) (see Fig.~\ref{fig:unc_thry}(b)).
The optimal \(\Gamma_t t_{RD}\) also decreases when the sampling interval \(\Gamma_t t_0\) decreases (see Fig.~\ref{fig:unc_thry}(c)).
For small \(\Gamma_t t_{RD}\), small \(\Gamma_t t_0\) reduces \(s_{\rm min}(\Gamma_t)\) compared to large \(\Gamma_t t_0\) by increasing the amount of averaging over the white noise before the thermomechanical noise becomes uncorrelated.
However, the lower \(s_{\rm min}(\Gamma_t)\) also makes ring-downs with small \(\Gamma_t t_0\) more sensitive to the decorrelation of the thermomechanical noise, which reduces the optimal \(\Gamma_t t_{RD}\).

The benefit of measuring for the optimal ring-down time depends on the relative amount of white noise and thermomechanical noise.
When \(N\) is large and \(\sigma_w\gg\sigma_{th}\), the optimal \(\Gamma_t t_{RD}\approx 4\) reduces \(u(\Gamma_t)\) by a factor of approximately \(1.3\) compared to \(\Gamma_t t_{RD}=10\) (see Fig.~\ref{fig:unc_thry}(a)).
When \(\sigma_{th}\gg \sigma_w\), measuring with the optimal \(\Gamma_t t_{RD}\) reduces \(u(\Gamma_t)\) by a factor larger than two compared to \(\Gamma_t t_{RD}=10\) (see pink curve in Fig.~\ref{fig:unc_thry}(a) and blue curve in Fig.~\ref{fig:unc_thry}(b)).
Changing the sampling rate causes the benefit of the optimal \(\Gamma_t t_{RD}\) relative to $\Gamma_t t_{RD}=10$ to vary by approximately \(20~\si{\percent}\) for the range of sampling rates shown in Fig.~\ref{fig:unc_thry}(c).

\section{Optimal Readout}
\label{sec:opt}

\begin{figure}
    \includegraphics[width=\columnwidth]{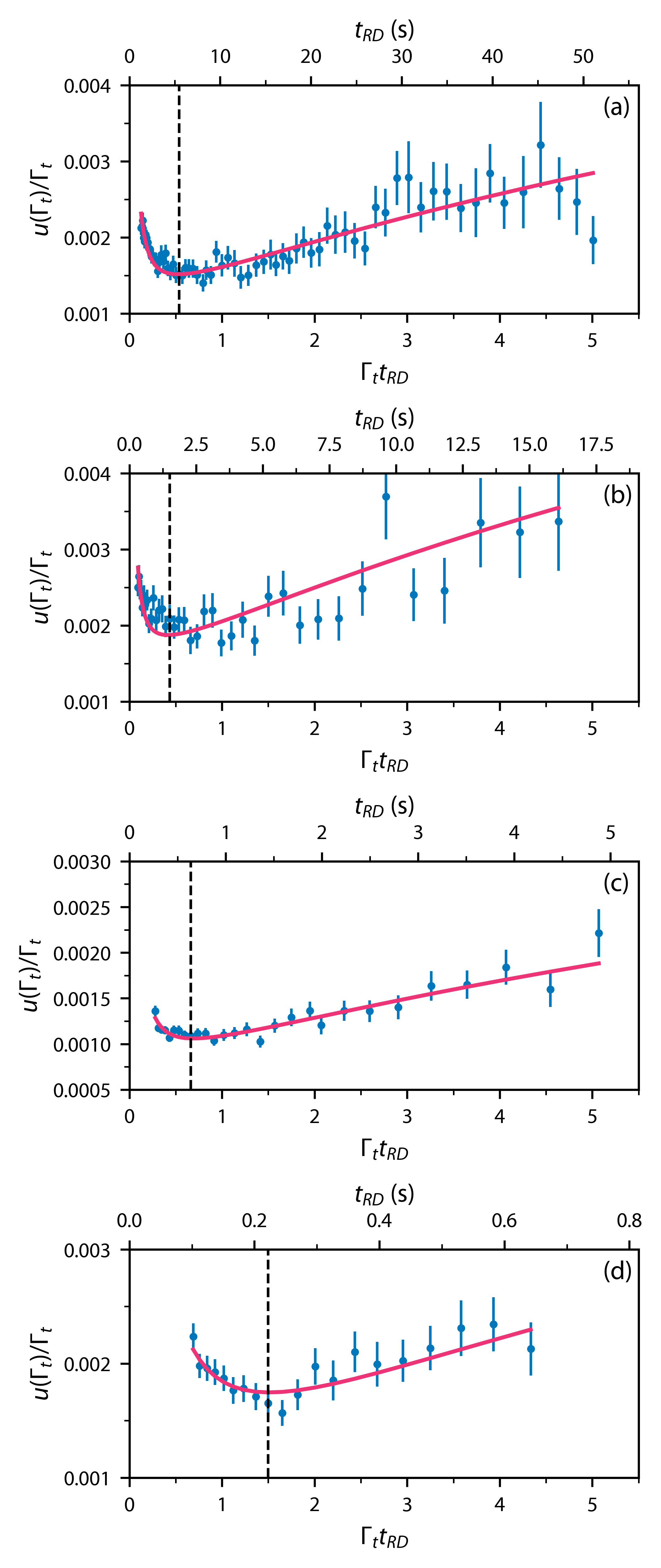}
    \caption{\label{fig:unc_meas}
    Fractional standard uncertainty in \(\Gamma_t\) as a function of individual ring-down time \(t_{RD}\) at fixed total measurement time \(\tau\).
    Blue circles show the measured \(u(\Gamma_t)/\Gamma_t\) at base pressure (\(2\times10^{-5}~\si{\pascal}\)) (a), \(7~\si{\milli\pascal}\) of argon (b), \(54~\si{\milli\pascal}\) of argon (c), and \(402~\si{\milli\pascal}\) of argon (d).
    Pink curves are fits to the data with the Cram\'er-Rao bound (Eq.~\eqref{eq:bound}).
    Dashed vertical lines show the optimal \(\Gamma_t t_{RD}\) extracted from each fit.
    Error bars show the approximate standard uncertainty of \(u(\Gamma_t)/\Gamma_t\). 
    }
\end{figure}

We experimentally determine the optimal individual ring-down time \(t_{RD}\) by measuring \(u(\Gamma_t)\) as a function of \(t_{RD}\).
Figure~\ref{fig:unc_meas}(a), Fig.~\ref{fig:unc_meas}(b), Fig.~\ref{fig:unc_meas}(c), and Fig.~\ref{fig:unc_meas}(d) show the measured \(u(\Gamma_t)/\Gamma_t\) with Cram\'er-Rao bound fits at base pressure (\(2\times10^{-5}~\si{\pascal}\), \(\Gamma_t=\Gamma_0\approx 0.097~\si{\per\second}\)), \(7~\si{\milli\pascal}\) of argon (\(\Gamma_t\approx 0.29~\si{\per\second}\)), \(54~\si{\milli\pascal}\) of argon (\(\Gamma_t\approx 1.0~\si{\per\second}\)), and \(402~\si{\milli\pascal}\) of argon (\(\Gamma_t\approx 6.8~\si{\per\second}\)), respectively.
Both the experimental data and fits account for the dead time \(t_d\) at each ring-down time \(t_{RD}\) (see Eq.~\eqref{eq:stand_unc}).
We vary \(\sigma_w/V_{\rm exc}\) and \(\sigma_{th}/V_{\rm exc}\) in the Cram\'er-Rao bound fits.
In \{Fig.~\ref{fig:unc_meas}(a), Fig.~\ref{fig:unc_meas}(b), Fig.~\ref{fig:unc_meas}(c), Fig.~\ref{fig:unc_meas}(d)\}, the fits report \(\sigma_w/V_{\rm exc}=\{1.1\times10^{-3},9\times10^{-4},1.1\times10^{-3},2.1\times10^{-3}\}\) and \(\sigma_{th}/V_{\rm exc}=\{1.4\times10^{-3},1.7\times10^{-3},9\times10^{-4},9\times10^{-4}\}\).
The fits reproduce the main features of the data, confirming that the Cram\'er-Rao bounds derived in Sec.~\ref{sec:cr} accurately describe the dominant noise sources.

 
From base pressure to \(54~\si{\milli\pascal}\) of argon, using the optimal \(\Gamma_t t_{RD}\) reduces \(u(\Gamma_t)\) by a factor of approximately \(1.8\) compared to using \(t_{RD}\approx 5/\Gamma_t\) (see Fig.~\ref{fig:unc_meas}(a), Fig.~\ref{fig:unc_meas}(b), and Fig.~\ref{fig:unc_meas}(c)).
The optimal \(\Gamma_t t_{RD}\) also exhibits minimal variation around \(\Gamma_t t_{RD}\approx 0.6\) at low pressure, corresponding to the fact that the optimal \(t_{RD}\) is inversely proportional to pressure in this range (see the upper horizontal axis in each subplot of Fig.~\ref{fig:unc_meas}).
The dependence of \(t_{RD}\) on pressure is expected given the physical operating principle of damping-based optomechanical pressure sensors.\cite{Christian1966, lubbe2011, reinhardt2024, Green2025, Cavalleri2010, Martinetz2018}
Figure~\ref{fig:unc_meas}(d) shows that the optimal \(\Gamma_t t_{RD}\) increases to \(\Gamma_t t_{RD}\approx1.5\) at \(402~\si{\milli\pascal}\) of argon.
The increase in optimal ring-down time occurs because the sampling interval \(\Gamma_t t_0\) is roughly four times larger in Fig.~\ref{fig:unc_meas}(d) than it is in the prior three subplots of Fig.~\ref{fig:unc_meas}.
The shift in the optimal individual ring-down time is consistent with the results of Fig.~\ref{fig:unc_thry}(c) in Sec.~\ref{sec:cr}.

The optimal \(t_{RD}\) for a particular optomechanical pressure sensor interacting with any gas at any pressure can be estimated from a measurement of the optimal \(t_{RD}\) at base pressure (\textit{e.g.} Fig.~\ref{fig:unc_meas}(a)).
As explored in Sec.~\ref{sec:cr}, the optimal \(t_{RD}\) depends on four parameters: \(\sigma^2_w\), \(\sigma^2_{th}\), \(t_0\), and \(\Gamma_t\).
The noise variances \(\sigma^2_w\) and \(\sigma^2_{th}\) are approximately independent of pressure because they are properties of the measurement system and optomechanical pressure sensor, respectively.
The total mechanical damping rate \(\Gamma_t\) is given by the sum of the optomechanical pressure sensor's intrinsic damping rate \(\Gamma_0\) (determined via ring-down measurement at base pressure, see Sec.~\ref{sec:procedure}) and the pressure-induced damping rate, which can be calculated following Refs.~\onlinecite{Cavalleri2010, Martinetz2018, Green2025}.
In the molecular flow regime -- where damping-based optomechanical pressure sensors are linear\cite{lubbe2011, reinhardt2024, Green2025} -- the pressure-induced damping rate is proportional to the pressure \(P\) and \(\sqrt{m_g}\), where \(m_g\) is the molecular mass of the gas.\cite{Christian1966, lubbe2011, reinhardt2024, Green2025, Cavalleri2010, Martinetz2018}
If the sampling interval \(t_0\) is adjusted as the gas pressure increases such that \(\Gamma_t t_0\) matches the base pressure measurement, then the optimal \(\Gamma_t t_{RD}\) will be independent of pressure (see Fig.~\ref{fig:unc_thry}(c)).
The optimal \(t_{RD}\) for a gas at pressure \(P\) can then be approximated by scaling the optimal \(t_{RD}\) at base pressure by \(\Gamma_0/\Gamma_t\).

The fractional standard uncertainties \(u(\Gamma_t)/\Gamma_t\) that we observe for \(t_{RD}\ll 1/\Gamma_t\) are comparable to or lower than \(u(\Gamma_t)/\Gamma_t\) achieved with \(t_{RD}\gg 1/\Gamma_t\) (see Fig.~\ref{fig:unc_meas}).
As a result, the pressure measurement rate can be increased significantly compared to the results of Refs.~\onlinecite{reinhardt2024, Green2025} without compromising precision.
Here, we have achieved pressure measurement rates up to approximately \(10~\si{\per\second}\) at approximately \(70~\si{\percent}\) duty cycle.
Our duty cycle is currently limited by the strength of the drive to the piezoelectric shaker and the acquisition of data during excitation to assess \(t_d\).
Our pressure measurement rate is currently limited to \(t_{RD}\gtrsim 0.05/\Gamma_t\) by \(V_{\rm exc}\) overshoot during switch turn-off.
An individual ring-down time of \(t_{RD}= 0.05/\Gamma_t\) corresponds to pressure measurement rates of approximately \(0.5~\si{\per\second}\) at base pressure and approximately \(100~\si{\per\second}\) at approximately \(400~\si{\milli\pascal}\) of Argon.
We note that the limitations on the measurement rate and duty cycle are purely technical and could be straightforwardly improved upon in an engineered commercial sensor.
For example, in a limited set of measurements using another optomechanical trampoline resonator, we have found that \(98~\si{\percent}\) duty cycle is achievable by setting \(a_d=0.01\) (see Eq.~\eqref{eq:dead}) and modestly increasing the piezoelectric shaker drive strength.

\begin{figure}
    \includegraphics[width=\columnwidth]{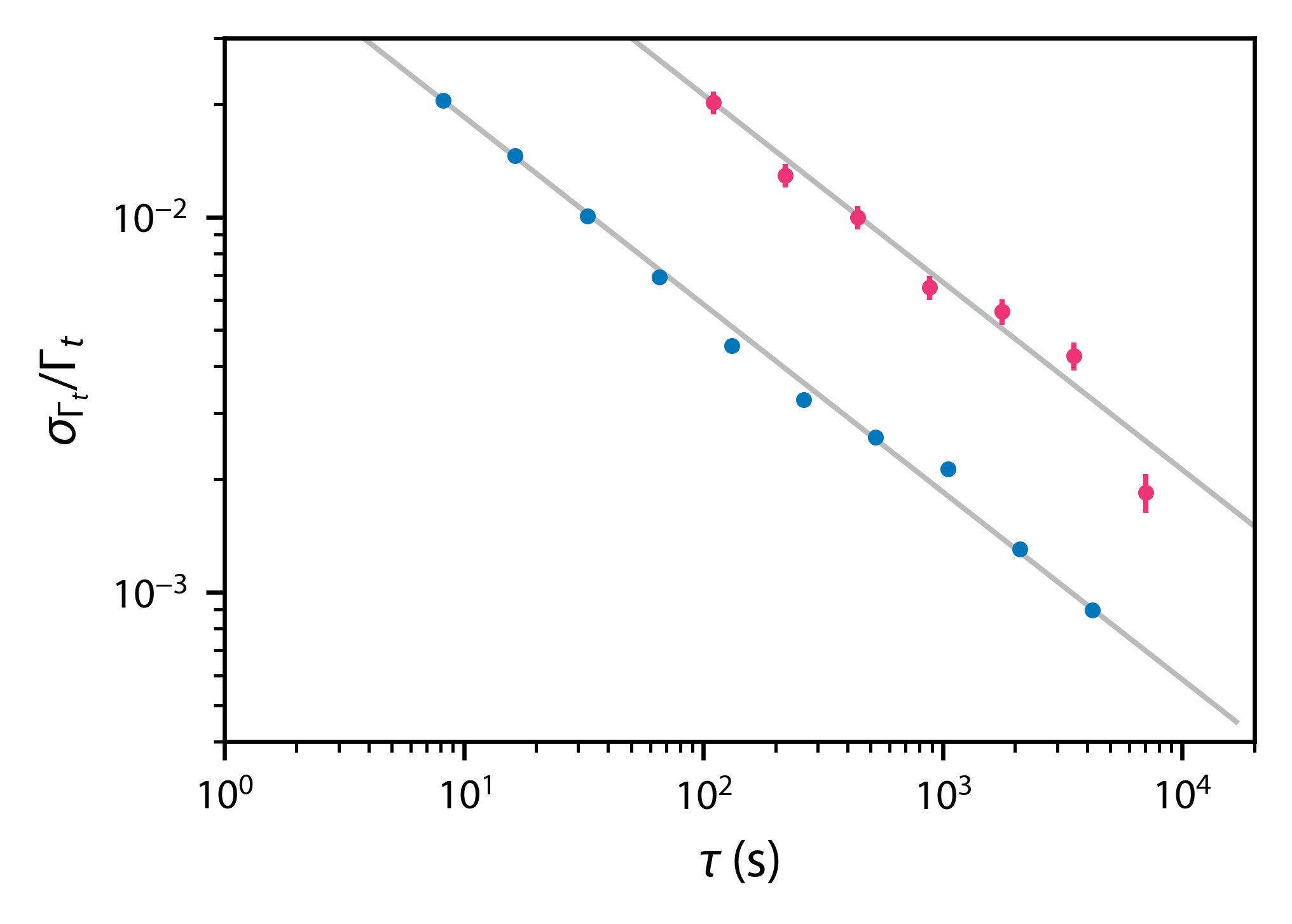}
    \caption{\label{fig:adevs}
    Fractional Allan deviation \(\sigma(\Gamma_t)/\Gamma_t\) as a function of total measurement time \(\tau\) at base pressure.
    The blue circles show \(\sigma(\Gamma_t)/\Gamma_t\) measured using the optimal ring-down time \(\Gamma_t t_{RD} \approx 0.7\).
    The pink circles show \(\sigma(\Gamma_t)/\Gamma_t\) measured using the long ring-down time \(\Gamma_t t_{RD} \approx 9.9\).
    Gray lines show a \(\tau^{-1/2}\) trend from the initial \(\sigma(\Gamma_t)/\Gamma_t\) for each Allan deviation.
    Error bars show the approximate statistical uncertainty of \(\sigma(\Gamma_t)/\Gamma_t\) (many error bars are smaller than the data points).
    }
\end{figure}

The benefit of using the optimal \(t_{RD}\) is more easily visualized using the Allan deviation \(\sigma(\Gamma_t)\).
Figure~\ref{fig:adevs} shows the fractional Allan deviation \(\sigma(\Gamma_t)/\Gamma_t\) at base pressure (\(\Gamma_t\approx 0.093~\si{\per\second}\)) for repeated ring-down measurements with approximately \(80~\si{\percent}\) duty cycle.
Using the optimal \(\Gamma_t t_{RD}\approx 0.7\) allows either a reduction of \(\sigma(\Gamma_t)\) by a factor of approximately three at fixed \(\tau\) or a reduction of \(\tau\) by a factor of approximately 10 at fixed \(\sigma(\Gamma_t)\) compared to \(\Gamma_t t_{RD}\approx 10\).
The reduction in \(\sigma(\Gamma_t)\) at base pressure and fixed \(\tau\) indicates that our measurements are dominated by thermomechanical noise from the trampoline.
The reduction in \(\sigma(\Gamma_t)\) is consistent with the \(u(\Gamma_t)\) measurements in Fig.~\ref{fig:unc_meas} and the results of Sec.~\ref{sec:cr}.

We observe that \(\sigma(\Gamma_t)/\Gamma_t\) decreases as \(\tau^{-1/2}\) for the duration of our measurements (\(\tau\approx 5\times10^3~\si{\second}\) for both data sets shown in Fig.~\ref{fig:adevs}).
The useful measurement time is so long because the effect of the thermomechanical noise on the fit to Eq.~\eqref{eq:quad} is reset at the beginning of each ring-down.
As a result, the uncertainties in individual ring-down measurements are uncorrelated.
At \(\tau=10^{3}~\si{\second}\) with \(\Gamma_t t_{RD}\approx 0.7\), \(\sigma(\Gamma_t)/\Gamma_t\approx0.2~\si{\percent}\), which corresponds to a resolvable pressure rise of \(17~\si{\micro\pascal}\)  of N\(_2\) at \(20~\si{\celsius}\).
This is approximately a factor of two reduction in resolvable pressure rise in approximately a factor of five less measurement time compared to our prior results.\cite{Green2025}

\section{Conclusion}
\label{sec:con}

We demonstrate and study a fast readout method for tethered optomechanical pressure sensors.
Our method allows measurement of the total mechanical damping rate \(\Gamma_t\) with a ring-down time \(t_{RD}<1/\Gamma_t\) and therefore allows pressure measurement rates that exceed \(\Gamma_t\).
We show, using both the Cram\'er-Rao bound and the experimentally determined uncertainty, that probing each ring-down for the optimal \(t_{RD}\) reduces the statistical uncertainty in \(\Gamma_t\) and \(P\).
Near base pressure, using the optimal \(t_{RD}\) (combined with the switch to lock-in detection) improves the resolvable pressure rise by approximately a factor of two compared to our prior work.\cite{Green2025}
Alternatively, the pressure measurement rate can be increased up to approximately \(20\Gamma_t\) with little reduction in precision compared to prior work.\cite{reinhardt2024, Green2025}
Our results show that tethered optomechanical pressure sensors can achieve pressure measurement rates comparable to levitated mechanical pressure sensors,\cite{Fremerey1985, Blakemore2020} removing a significant obstacle to their adoption in industrial settings.
We also anticipate that our Cram\'er-Rao noise models may aid development of optimal readout schemes for other sensors that measure exponential decays, such as cold-atom vacuum standards.\cite{Ehinger2022, Barker2022}

\section*{Acknowledgments}

We thank J. Manley and S. Eckel for their careful reading of the manuscript.
We also thank A. Migdall and I. Spielman for loaning us equipment used in this work.

\section*{Data Availability}

The data that support the findings of this study are available from the corresponding author upon reasonable request.

\section*{Conflict of Interest}

SRC and DSB are have filed US provisional patent application 64/146,720.



\appendix
\section{\label{sec:psd}Autocorrelation and power spectral density of demodulated thermomechanical noise}

The noise present on the demodulated signal~$X(t)$ exiting the lock-in amplifier contributes to the uncertainty in our determination of the decay rate $\Gamma_t$.
Here, we take the noise $q(t)$ entering the lock-in amplifier to be of thermal origin, described by the autocorrelation $R_{qq}(\delta t)=E\{q(t)q(t-\delta t)\}$, where \(\delta t\) is the time difference and the notation $E\{\}$ denotes an ensemble average.
The autocorrelation $R_{qq}(\delta t)$ is given by the Fourier transform of the well-known power spectrum for the thermally driven harmonic oscillator with natural frequency~$\omega_0$ and damping rate $\Gamma_t$.\cite{Hauer2015,Papoulis}
In the limit of a low damping rate~$\Gamma_t$, it is\cite{Papoulis}
\begin{equation}
    R_{qq}(\delta t)=\frac{k_BT}{m_{\rm eff}\omega_0^2}e^{-\Gamma_t|\delta t|/2}\cos\omega_0\delta t,
    \label{eqn: Rqq}
\end{equation}
where \(m_{\rm eff}\) is the effective mass, \(T\) is the temperature, and \(k_B\) is the Boltzmann constant.
For the pressure range that we explore in Sec.~\ref{sec:opt}, $\Gamma_t\ll\omega_0\approx 2\pi\times30~\si{\kilo\hertz}$, so the low damping limit is always satisfied.
Our goal now is to determine the autocorrelation of the demodulated thermomechanical noise~$X_q(t)$ exiting the lock-in amplifier.

We start by expressing the random variable~$q(t)$ as
\begin{equation}
    q(t)=\frac{1}{2}\left(q_0(t)e^{i\omega_0t}+q_0^*(t)e^{-i\omega_0t}\right).
    \label{eqn: SVArep}
\end{equation}
The fact that $R_{qq}$ is a function of only the time difference~$\delta t$ places some important constraints on the statistical properties of the complex amplitude~$q_0(t)$.
Indeed, substituting the representation~\eqref{eqn: SVArep} into the definition of the autocorrelation gives
\begin{eqnarray}
    R_{qq}(\delta t)&=&E\{q(t)q(t-\delta t)\} \nonumber \\
    &=&\frac{1}{4}\left(E\{q_0(t)q_0(t-\delta t)\}e^{2i\omega_0t}e^{-i\omega_0\delta t} \right. \nonumber \\
    & &\left.+E\{q_0(t)q_0^*(t-\delta t)\}e^{i\omega_0\delta t}+c.c.\right) 
    \label{eqn: Rqq1}
\end{eqnarray}
where $c.c.$ denotes the complex conjugate.
Since $R_{qq}$ is known from Eq.~(\ref{eqn: Rqq}) to be a function of only~$\delta t$, it must be the case that
\begin{equation}
E\{q_0(t)q_0(t-\delta t)\}=0
\label{eqn: constraint}
\end{equation}
for all~$t$.
It then follows that
\begin{equation}
R_{qq}(\delta t)=\frac{1}{2}\text{Re}\left(E\{q_0(t)q_0^*(t-\delta t)\}e^{i\omega_0\delta t}\right) 
\label{eqn: Rqq2}
\end{equation}
and comparison with~(\ref{eqn: Rqq}) yields
\begin{equation}
\text{Re}\left(E\{q_0(t)q_0^*(t-\delta t)\}\right) =\frac{2k_BT}{m_{\rm eff}\omega_0^2}e^{-\Gamma_t|\delta t|/2}.
\label{eqn: cuteresult}
\end{equation}

Returning to the operation of the lock-in amplifier, we start by scaling the thermomechanical noise~$q(t)$ by the same constant ${\cal{C}}$ implicit in Eq.~(\ref{eq:quad}) relating the signal voltage~$X(t)$ to the trampoline's amplitude of motion \(A(t)\), to obtain a  thermal noise signal voltage~${\cal{C}}q(t)$.  The demodulated thermomechanical  noise~$X_q(t)$ is obtained by 
multiplying~${\cal{C}}q(t)$ by a local oscillator with angular frequency $\omega_{LO}$ and passing the output through a low-pass filter (see Fig.~\ref{fig:apparatus}).
Denoting the action of the low-pass filter by $\langle\,\rangle$, taking the relevant case $\omega_{LO}=\omega_0$, and using~(\ref{eqn: SVArep}), we have
\begin{eqnarray}
X_q(t)&=&{\cal{C}}\langle q(t)\cos\omega_{LO}t\rangle \nonumber \\
&=&\frac{{\cal{C}}}{2}\langle \left(q_0(t)e^{i\omega_0t}+q_0^*(t)e^{-i\omega_0t}\right)\cos\omega_0t\rangle \nonumber \\
&=&\frac{{\cal{C}}}{4} \left(q_0(t)+q_0^*(t)\right).
\label{eqn: Xoft}
\end{eqnarray}
The autocorrelation of~$X_q(t)$ is then given by
\begin{eqnarray}
R_{X_qX_q}(\delta t)&=&\frac{{\cal{C}}^2}{16} E\{\left(q_0(t)+q_0^*(t)\right) \left(q_0(t-\delta t)+q_0^*(t-\delta t)\right)\} \nonumber \\
&=&\frac{{\cal{C}}^2}{8}\text{Re}\left(E\{q_0(t) q_0(t-\delta t)+q_0(t)q_0^*(t-\delta t)\}\right) \nonumber \\
&=&\frac{{\cal{C}}^2}{4}\frac{k_BT}{m_{\rm eff}\omega_0^2}e^{-\Gamma_t|\delta t|/2},
\label{eqn: RXX1}
\end{eqnarray}
where we have used~(\ref{eqn: constraint}) and~(\ref{eqn: cuteresult}).
Taking the Fourier transform, we obtain the power spectral density of the thermomechanical noise output by the lock-in amplifier:
\begin{equation}
S_{X_qX_q}(\omega)=\frac{{\cal{C}}^2k_B T}{4m_{\rm eff}\omega_0^2}\frac{\Gamma_t }{\omega^2+\frac{\Gamma_t^2}{4}}, 
\label{eqn:Sxx2}
\end{equation}
where \(\omega=2\pi f\) is the angular frequency.
The expressions~\eqref{eqn: RXX1} and~\eqref{eqn:Sxx2} are related to the variance of the thermal noise~$\sigma_{th}^2$ used in the main text by
\begin{equation}
\label{eq:sigma_th}
\sigma_{th}^2= R_{X_qX_q}(0)=\frac{{\cal{C}}^2}{4}\frac{k_BT}{m_{\rm eff}\omega_0^2},
\end{equation}
so that
\begin{equation}
R_{X_qX_q}(\delta t)=\sigma_{th}^2 e^{-\Gamma_t|\delta t|/2}
\end{equation}
and
\begin{equation}
S_{X_qX_q}(\omega)=\sigma_{th}^2\frac{\Gamma_t }{\omega^2+\frac{\Gamma_t^2}{4}}. 
\end{equation}
For frequencies $\omega>>\Gamma_t/2$, the demodulated thermomechanical noise spectrum has the $\omega^{-2}$ dependence discussed in Sec.~\ref{sec:procedure}.

\section{\label{sec:covar}Covariance matrix}
Individual measurements of the random variable~$X_j$ are given by
\begin{equation}
X_j=\overline{X}_j+\delta X(j t_0),
\end{equation}
where the mean is given by Eq.~(\ref{eq:rand}) and $\delta X(j t_0)$ describes the noise measured at time~$j t_0$. The covariance matrix is
\begin{eqnarray}
C_{jk}&=&E\{X_jX_k\}-\overline{X}_j\overline{X}_k \nonumber \\
&=&E\{\delta X(j t_0)\delta X(k t_0)\}, 
\end{eqnarray}
since the cross-terms in the expansion of $E\{X_jX_k\}$ vanish.
In this work, we consider independent thermal and white noise sources $\delta X_{th}$ and $\delta X_{w}$; writing $\delta X=\delta X_{th}+\delta X_{w}$, we find
\begin{eqnarray}
C_{jk}&=&E\{\delta X_{th}(j t_0)\delta X_{th}(k t_0)+\delta X_{w}(j t_0)\delta X_{w}(k t_0)\} \nonumber \\
&=&R_{th}((j-k) t_0)+R_{w}((j-k) t_0) \nonumber \\
&=&\sigma_{th}^2 e^{-\Gamma_tt_0|j-k |/2}+\sigma_{w}^2\delta_{jk}.
\end{eqnarray}

\bibliography{fast_readout.bib}

\end{document}